\documentclass[twocolumn,showpacs,preprintnumbers,amsmath,amssymb,floatfix,superscriptaddress]{revtex4-1}

\usepackage{graphicx}
\usepackage{amsmath,amsfonts,amssymb}
\usepackage{graphicx}
\usepackage{color}
\usepackage{bbold}

\usepackage[normalem]{ulem}
\usepackage{todonotes}
\usepackage[bookmarks,
            breaklinks,
            pdfstartview=FitH, 
            pdftitle={},
            pdfauthor={},
            pdfkeywords={}
            ]{hyperref}
\usepackage{ulem}
\hypersetup{
  colorlinks   = true,  
  urlcolor     = blue,  
  linkcolor    = blue,  
  citecolor    = red    
}

\usepackage[margin=1.in]{geometry}
\usepackage{graphicx}
\usepackage{amsmath}
\usepackage{amssymb}
\usepackage{mathtools}
\usepackage{amsthm}
\usepackage{tikz-cd}
\usepackage{subfigure}

\begin{document}

\title{High-precision analysis of Feshbach resonances in a Mott insulator}
\author{Thomas Secker}
\affiliation{Eindhoven University of Technology, P.~O.~Box 513, 5600 MB Eindhoven, The Netherlands}
\author{Jesse Amato-Grill}
\affiliation{Research Laboratory of Electronics, MIT-Harvard Center for Ultracold Atoms, and Department of Physics, Massachusetts Institute of Technology, Cambridge, Massachusetts 02139, USA}
\affiliation{Department of Physics, Harvard University, Cambridge, Massachusetts 02138, USA}
\author{Wolfgang Ketterle}
\affiliation{Research Laboratory of Electronics, MIT-Harvard Center for Ultracold Atoms, and Department of Physics, Massachusetts Institute of Technology, Cambridge, Massachusetts 02139, USA}
\author{Servaas Kokkelmans}
\affiliation{Eindhoven University of Technology, P.~O.~Box 513, 5600 MB Eindhoven, The Netherlands}

\begin{abstract}

We show that recent high-precision measurements of relative on-site interaction energies $\Delta U$ in a Mott insulator require a theoretical description beyond the standard Hubbard-model interpretation, when combined with an accurate coupled-channels calculation.
In contrast to more sophisticated lattice models, which can be elaborate especially for parameter optimization searches, we introduce an easy to use effective description of $U$ valid over a wide range of interaction strengths modeling atomic pairs confined to single lattice sites. 
This concise model allows for a straightforward combination with a coupled-channels analysis.
With this model we perform such a coupled-channels analysis of high-precision $^7$Li spectroscopic data on the on-site interaction energy $U$, which spans over four Feshbach resonances and provide
an accurate and consistent determination of  the associated resonance positions.
Earlier experiments on three of the Feshbach resonances are consistent with this new analysis.
Moreover, we verify our model with a more rigorous numerical treatment of the two atom system in an optical lattice.

\end{abstract}

\maketitle

\section{Introduction}

Precise knowledge and control of the atomic interactions is essential to a wide range of modern ultracold atom experiments.
Coupled-channels models can accurately describe ultracold collision properties \cite{Taylor72} of a two-atom system by detailed interaction potentials that are finetuned by just a few parameters, to match atom loss spectra \cite{Marte:2002}, photoassociation rates and collisional cross sections \cite{Kempen:2002}, interference patterns between s-wave and d-wave collisions \cite{Volz:2005}, and rf molecule association data \cite{Gross:2011}.
Accurate information on the interaction strength close to Feshbach resonances \cite{Feshbach92} is needed to determine the three-body parameter of a strongly-interacting Bose gas \cite{Gross:2010,Chapurin:2019}. Also the collisional energy-dependence can be substantial, when operating a cesium microgravity clock \cite{Bennett:2017}, or for the determination of the nature of a spinor condensate \cite{Kawaguchi:2012,Stamper-Kurn:2013}.
Intermediate descriptions are often required to embed the collisional properties of two atoms into the macroscopic environment of the ultracold gas, where atoms may be held by magnetic or optical traps~\cite{Busch:1998,Schneider:2011}, or by optical lattices~\cite{Buchler:2010}. 
Given the collisional properties of two atoms one can then correctly account for collisional energy shifts, however, the collisional properties themselves depend on those energy shifts via the relative kinetic energy of the colliding atoms that is linked to the total energy of the two particles, which is shifted by the collisional energy shift.
Therefore a self-consistent approach is needed and the precision of the coupled-channels interaction parameters depends crucially on the precision of the intermediate descriptions. Using a less precise model may result in inconsistencies with respect to the two-body parameters when comparing between different experiments, or even when comparing data within a single experiment.

In this paper, we want to utilize a coupled-channels model in combination with an accurate but easy to use theoretical lattice model description of the on-site interaction energy, to find a consistent description of high-precision spectroscopic data of a two component Mott insulator with two particles per site, where the on-site energy was varied by using Feshbach resonances~\cite{Amato-Grill:2019}.
In this experiment, atoms in different spin states were confined in a cubic optical lattice geometry. 
Such systems can implement a spin-Heisenberg model \cite{Altman:2003}, which is of interest to study quantum magnetism. 
In the Mott insulating phase with doubly occupied sites individual atom pairs localize around each lattice site~\cite{Jaksch:1998,Bloch:2008}. 
At the same time the interactions between different spin components of the atoms can be controlled utilizing Feshbach resonances.
This condition results in an ideal situation to study the isolated two particle system.
Then, the strength of the atomic interaction characterized by the scattering length $a_S$ can be linked to the energy per localized atom pair $E(a_S)$ and thus to the on-site interaction energy $U$ defined as $U = E(a_S) - E(0)$. 
In Ref.~\cite{Amato-Grill:2019} high precision data on the on-site interaction energy has been acquired in such a two component Mott insulator using interaction spectroscopic techniques.

We show that the standard Hubbard model approximation~\cite{Jaksch:1998} leading to a linear relation between the on-site interaction energy $U$ and the scattering length $a_S$ is not accurate enough to describe the experimental data to the required precision even for moderate scattering lengths.
Instead we offer a simple effective description to approximate the effects of a cubic lattice potential for two identical particles of mass $m$, in which the on-site interaction energy $U$ is parametrized by a harmonic oscillator model with contact interactions~\cite{Busch:1998}. The harmonic oscillator frequency $\omega_\text{eff}$ should be adjusted to best represent the on-site potential. Note that this is not necessarily the frequency related to the second order approximation around a lattice potential minimum. We choose $\omega_\text{eff}$, such that the harmonic model matches the linear description from the Hubbard model close to the non-interacting case, where the Hubbard model should give the correct description up to first order.
The effective trapping frequency $\omega_\text{eff}$ is then completely determined by the non-interacting lattice parameters
\begin{equation} \label{eq:linearmatching}
\hbar \omega_\text{eff} = 2 \pi E_\text{r} \left( \frac{1}{k_\text{r}} \int |w_0(r)|^4 \text{d} r \right)^2 \, ,
\end{equation}
where $w_0$ is the lowest band Wannier function of the one dimensional single particle system, $E_\text{r} = \hbar^2 k_\text{r}^2 / (2 m)$ is the recoil energy with $k_\text{r} = 2 \pi / \lambda$ and $\lambda$ is the wavelength of the lattice light.
All in all our effective model is meant to describe the two particle system with contact interactions including the full lattice potential. This system has been theoretically studied in detail~\cite{Buchler:2010,Deuretzbacher:2008}.
To make the connection to a more complete description of the inter-particle interactions, the strength of the contact interaction is gauged to match the free scattering properties of the full coupled-channels system for a fixed collision energy $\epsilon$.
In ultra-cold experiments it is often sufficient to take $\epsilon \approx 0$ and thus to match the scattering length $a_S$.
To achieve a more accurate description we want to match the full coupled-channels and contact system at the correct relative collision energies $\epsilon$ in a self-consistent manner.
We find an approximate expression $\epsilon(a_S) \approx E(a_S) - \mathcal{E}_0$, which is given by the total energy per localized atom pair depending on the scattering length minus the energy in the center of mass (c.m.) direction.

By gauging the contact interaction strength and matching the collision energy we can thus find the on-site interaction energy $U$ for a full coupled-channels description in a lattice environment. 
To check the validity and limitations of the effective description we compare to the full contact system in the lattice using a direct diagonalization approach as presented in~\cite{Deuretzbacher:2008,Grishkevich:2011}. In addition we investigate the collision energies from the full contact bound state wave functions to match the contact to the coupled-channels solution at the correct relative energies. This analysis also enables us to identify regimes in which the simple contact approximation is likely to fail, for example close to narrow resonances, where the full systems scattering properties show strong dependence on the relative collision energy. 
Finally we apply our approach to perform a full coupled-channels analysis of recent experimental data on $^7$Li~\citep{Amato-Grill:2019}.

\section{theory}

We start with some remarks on the system we aim to model. All in all we have a many-body system in mind, but since we assume the system to be in a Mott insulating phase with two atoms per lattice site in a regime where the interaction with atoms at neighboring sites may be neglected, we restrict to a simple two particle model system with both particles well localized at a single lattice site.
The two particle system in a lattice is described by the following Hamiltonian
\begin{equation}
H = \frac{\mathbf{p}_1^2}{2 m} + \frac{\mathbf{p}_2^2}{2 m} +  V_\text{opt}(\mathbf{r}_1,\mathbf{r}_2) + V_\text{int}(\mathbf{r}_1-\mathbf{r}_2) \, ,
\end{equation}
with $\mathbf{p}_i$ and $\mathbf{r}_i$ the momentum and position operators of particle $i \in \{ 1 ,2 \}$, $m$ the mass of the particles, $V_\text{int}$ the interaction between the particles and $V_\text{opt}$ the lattice potential.

To simplify the system further one can mimic the interactions between the particles with a contact interaction of variable interaction strength. By adjusting this interaction strength we can correctly represent the physics at length scales bigger than the range $r_0$ of the real interaction potential. 
Note however that this approximation will only be good around some fixed relative energy $\epsilon$ between the particles. 

In the following we give a detailed analysis of the contact interaction case and focus in the second part on the connection of contact to coupled-channels model. We include a discussion of effects related to the non-conserved relative collision energy in the lattice scenario.

\subsection{Contact interaction}

To mimic the interactions between the particles we introduce a contact interaction of variable interaction strength, implemented with a Bethe-Peierls boundary condition~\cite{Bethe:1935}. We parametrize the interaction strength by the scattering length $a_S$ of the interaction such that we have
\begin{equation}
V_\text{int} (\mathbf{r}_1 - \mathbf{r}_2) = \frac{4 \pi \hbar^2 a_S}{m} \delta(\mathbf{r}_1-\mathbf{r}_2) \, .
\end{equation}
Let us first consider the case of a weakly interacting system where $a_S \rightarrow 0$ and a perturbative analysis is possible. We closely follow the derivation of the Hubbard model and thus also make the connection to the many body theory. We start from the non interacting case $(a_S = 0)$, such that we can restrict to the single particle scenario. We consider a cubic optical lattice
\begin{align}
V_\text{opt} (\mathbf{r}_1,\mathbf{r}_2) & = \sum_{i \in \{ 1,2 \}} V_0 \, E_\text{r} \left[ \text{sin}(k_\text{r} x_i)^2 \right. \nonumber\\
& \phantom{=} \left. \qquad + \,  \text{sin}(k_\text{r} y_i)^2 +\text{sin}(k_\text{r} z_i)^2 \right] \, ,
\end{align}
with $V_0$ the lattice depth in recoil energies $E_\text{r}$. In this case the band structure as well as the single particle wave functions are known and can be expressed in terms of solution to Mathieu's equation \cite{Bloch:2008}. 
However, the so called Bloch solutions to the problem are not localized, but one can combine the Bloch waves of each band $n$ to find the band's Wannier function $\mathbf{w}_n (\mathbf{r}_j)$~\cite{Kohn:1959}.
The Wannier function is real valued and well localized around $\mathbf{r}_j = 0$ for deep lattices $E_\text{r} \rightarrow \infty$.
Together with versions $\mathbf{w}_n (\mathbf{r}_j - \mathcal{R}_i)$ shifted by a lattice vector $\mathcal{R}_i$ the Wannier function provides a complete basis set for the $n$-th band subspace. In addition the Wannier functions $\mathbf{w}_n$ approach eigenfunction solutions of a harmonic approximation around the lattice sites in the deep lattice limit and they are approximate solutions to the Schr\"odinger equation up to the width of the respective bands. To approximate the system's low energy part around a single lattice site it is therefore reasonable to project onto the  lowest band Wannier function $\mathbf{w}_0 (\mathbf{r}_j)$ 
such that the first order correction in energy will be 
\begin{equation}
U = \langle W | V_\text{int} | W \rangle \, ,
\end{equation}
where $W = \mathbf{w}_0(\mathbf{r}_1)\mathbf{w}_0(\mathbf{r}_2)$.
For the contact interaction we therefore obtain 
\begin{equation} \label{eq:linconv}
U = \frac{4 \pi \hbar^2 a_S}{m} \int |\mathbf{w}_0(\mathbf{r})|^4 \text{d} \mathbf{r} \, ,
\end{equation}
which leads to a linear correspondence between the energy correction $U$ and the scattering length $a_S$. Note that this corresponds exactly to the on-site interaction energy in the Hubbard model. Further we note that this treatment assumes the two-body wave function and therefore also the atomic density around a given lattice site to stay unaltered. To include the deformation of the wave function or atomic density caused by the interaction potential a higher order treatment is necessary.

To go beyond the linear regime Eq.~(\ref{eq:linconv}) we can approximate around a single lattice site with an interacting harmonic model \cite{Busch:1998}
\begin{align}
H_0(\omega_\text{eff}) &= \frac{\mathbf{p}_1^2}{2 m} + \frac{\mathbf{p}_2^2}{2 m} + \frac{4 \pi \hbar^2 a_S}{m} \delta(\mathbf{r}_1-\mathbf{r}_2) \nonumber\\
&\phantom{=} + \frac{1}{2} m \, \omega^2_\text{eff} \, \mathbf{r}_1^2 + \frac{1}{2} m \,  \omega^2_\text{eff} \, \mathbf{r}_2^2  \, .
\end{align}
The harmonic oscillator frequency $\omega_\text{eff}$ should be adjusted to best represent the lattice potential. Note that this is not necessarily the frequency related to the second order approximation around a lattice potential minimum.
We rather adjust the harmonic oscillator frequency such that it matches up with the first order result from the Hubbard model close to vanishing interaction strength (see Eq.~(\ref{eq:linearmatching}).
The spectrum of this effective harmonic model can be determined analytically and the shift in ground state energy $U$ is determined by~\cite{Busch:1998}
\begin{equation} \label{eq:ueffharm}
\frac{\sqrt{2} \,  \Gamma \left(\frac{U}{2  E_\text{HO}} \right)}{\ell_\text{HO} \Gamma \left(\frac{U}{2  E_\text{HO}} - \frac{1}{2}\right)} = \frac{1}{a_S} \, ,
\end{equation}
where $\Gamma$ denotes the gamma function, $E_\text{HO} = \hbar \omega_\text{eff}$ and $\ell_\text{HO} = \sqrt{\hbar / (m \omega_\text{eff})}$. Since the solutions to the harmonic model Hamiltonian are exact it naturally includes also the deformation of the wave function caused by the interaction, which manifests itself in the radial s-wave component of the relative wave function $\psi_{n,\ell=0} (r)$ given explicitly below in Eq.~(\ref{eq:relativeswavewavefunction}).

To analyze the accuracy and limitations of such an effective harmonic approach we calculate the spectrum of the contact system described by $H$ using a direct diagonalization approach similar to~\cite{Grishkevich:2011,Deuretzbacher:2008,Mentink:2009} making use of the eigenstates of the harmonic model. 
Such a numerical approach is accurate only around a single lattice site, therefore the population of neighboring lattice sites might be underestimated. 
This can lead to slight deviations between the numerical and Hubbard model description of the system, which we will discuss later.

We start by singling out the harmonic Hamiltonian in the full system by adding and substracting the harmonic potential
\begin{align}
H 
& = H_0(\omega_\text{eff}) +  V_\text{opt}(\mathbf{r}_1,\mathbf{r}_2) \nonumber \\ 
& \phantom{= H_0(\omega_\text{eff})}  - \frac{1}{2} m \omega^2_\text{eff} \left( r_1^2 + r_2^2 \right)\\
& = H_0(\omega_\text{eff}) + V_\Delta \, .
\end{align}
We will project on the eigenbasis of $H_0 (\omega_\text{eff})$ therefore the major task is to determine the coupling matrix elements resulting from the deviation from the lattice potential $V_\Delta$.
But before analyzing those in more detail let us first change to relative $\mathbf{r}$ and c.m. coordinates $\mathbf{R}$ with  
\begin{equation}
 \mathbf{r}_1 = \frac{1}{\sqrt{2}} \left( \mathbf{R} - \mathbf{r} \right)  \quad \text{and} \quad \mathbf{r}_2 = \frac{1}{\sqrt{2}} \left( \mathbf{R} + \mathbf{r} \right) \, .
\end{equation}
In addition we introduce units natural to the  harmonic model system $H_0(\omega_\text{eff})$, so all energies will be given in multiples of $\hbar \omega_\text{eff}$ and all lengths in multiples of $\sqrt{\hbar / (m \omega_\text{eff})}$.
The difference in potentials $V_\Delta$ separates into $x$, $y$ and $z$-direction
\begin{align}
V_\Delta
&= v_\Delta (X,x) + v_\Delta (Y,y)+v_\Delta (Z,z) \, ,
\end{align}
where the components $v_\Delta$ are given by
\begin{widetext}
\begin{align}
& v_\Delta (X,x) \nonumber \\
&= \frac{V_0}{2\sqrt{V_\text{eff}}} \left[
\text{sin}^2 \left(\frac{X - x }{ \sqrt{2} V_\text{eff}^{1/4}} \right) + \text{sin}^2 \left(\frac{X + x }{ \sqrt{2} V_\text{eff}^{1/4}} \right)\right] - \frac{1}{2} x^2 - \frac{1}{2} X^2\\
&= \frac{V_0}{\sqrt{V_\text{eff}}} \left[
\text{sin}^2\left(\frac{X}{ \sqrt{2} V_\text{eff}^{1/4}} \right) 
\text{cos}^2\left(\frac{x}{ \sqrt{2} V_\text{eff}^{1/4}} \right)
+ \text{cos}^2\left(\frac{X}{ \sqrt{2} V_\text{eff}^{1/4}} \right) 
\text{sin}^2\left(\frac{x}{ \sqrt{2} V_\text{eff}^{1/4}} \right)\right] 
- \frac{1}{2} x^2 - \frac{1}{2} X^2\\
&= \sum_{i,j} \alpha_{ij}(V_0,V_\text{eff}) x^{2 i}X^{2 j} \, .
\end{align}
\end{widetext}
In the last step we performed a taylor series  expansion and we have introduced the effective lattice depth parameter
$V_\text{eff} = \hbar^2 \omega_\text{eff}^2 / (2 E_\text{r})^2$. 
Note that in the series expansion only even powers of $x$ as well as $X$ occur. 
This leads to the symmetry properties of the system that have been discussed in a more general setting in~\cite{Grishkevich:2011}.

We now change to a basis of eigenstates $|N_X, N_Y, N_Z, n, \ell, m \rangle $ of the effective harmonic system. 
The $N_\circ $ are integers labeling the Harmonic oscillator eigenstates in the respective c.m. direction, whereas $n, \ell, m $ are quantum numbers in the relative direction, with $\ell$ the angular momentum quantum number, $m$ the magnetic quantum number and $n$ labels the solutions in the relative separation $r$. 
For $\ell \neq 0$ the quantum numbers $n, \ell, m $ just describe the usual non-interacting harmonic oscillator states, while for $\ell = 0$, $n$ labels the solutions of the harmonic model in the relative separation $r$. The general solution with correct behavior for $r \rightarrow \infty$ and relative energy $\epsilon_\text{rel}(n)$ is given up to a normalizing constant by
\begin{equation} \label{eq:relativeswavewavefunction}
\psi_{n,\ell=0} (r) \propto e^{-\frac{r^2}{2}} r  \text{U} \left(\frac{3}{4}-\frac{\epsilon _{\text{rel}}(n)}{2},\frac{3}{2},r^2\right) \, ,
\end{equation}
where $\text{U}$ is Tricomi's confluent hypergeometric function.
The quantization condition determining $\epsilon_\text{rel}(n)$ is then given by the boundary condition at $r=0$ and can be expressed in terms of the scattering length $a_S$
\begin{equation}
\frac{\sqrt{2} \Gamma \left(\frac{1}{4} \left(3-2 \epsilon _{\text{rel}}(n)\right)\right)}{\Gamma \left(\frac{1}{4}-\frac{\epsilon _{\text{rel}}(n)}{2}\right)} = \frac{1}{a_S} \, .
\end{equation}
We can now give the Hamiltonian $H_0 (\omega_\text{eff})$ in its diagonal form 
\begin{align}
& H_0(\omega_\text{eff}) \nonumber \\
& = \sum_{N_X, N_Y, N_Z, n, \ell, m} \bigg[ E(N_X, N_Y, N_Z, n, \ell, m)  \nonumber \\
& \phantom{=} |N_X, N_Y, N_Z, n, \ell, m \rangle \langle N_X, N_Y, N_Z, n, \ell, m  | \bigg] \, ,
\end{align}
where eigenenergies corresponding to the states $|N_X, N_Y, N_Z, n, \ell, m \rangle $
are given by 
\begin{align}
&E(N_X, N_Y, N_Z, n, \ell, m) \nonumber \\
&= \hbar \omega_\text{eff} \left[ 3 + N_X + N_Y + N_Z \right. \nonumber \\
& \phantom{=} \left. + \ell + \delta_{\ell,0} \epsilon_\text{rel}(n) + (1 - \delta_{\ell,0}) 2 n \right] \, .
\end{align}
To reduce the states we project on, we can use the symmetry properties of the Hamiltonian as we mentioned earlier. 
We have reflection symmetries in $X \rightarrow -X$, $Y \rightarrow -Y$ and $Z \rightarrow -Z$, therefore there will be just coupling between even or odd values of $N_X$,  $N_Y$ and $N_Z$ respectively. 
We also have the symmetry of inversion of $\mathbf{r} \rightarrow -\mathbf{r}$ equivalent to particle exchange, which leads to separation between even and odd values of $\ell$. For bosons we obviously need the even $\ell$ values. 
The symmetry under $(x,y) \rightarrow (-x,-y)$ leads to the restriction to even or odd values in $m$. Finally we could also change to a base 
\begin{align}
&| N_X, N_Y, N_Z, n, \ell, |m| \rangle_{S/A} \nonumber \\
& := \frac{1}{\sqrt{2 + 2\delta_{m,0}} } \left(|N_X, N_Y, N_Z, n, \ell, +|m| \rangle \right. \nonumber \\
& \phantom{:=} \left. \pm |N_X, N_Y, N_Z, n, \ell, -|m| \rangle \right) 
\end{align}
of symmetric and antisymmetric combinations in the sign of $m$.
The reflection symmetry in $z \rightarrow -z$ guaranties the separation in $S/A$.
We are interested in the component with $N_X$, $N_Y$, $N_Z$, $\ell$ and $|m|$ even and symmetric combinations ($S$) in $\pm |m|$, which corresponds to the ground state of the interacting harmonic model $H_0 (\omega_\text{eff})$ and thus to our solution up to zeroth order in $V_\Delta$ the corrections introduced by the full lattice potential.

\begin{figure}[htb!] 
	  \centering
	  \includegraphics[width = .5 \textwidth]{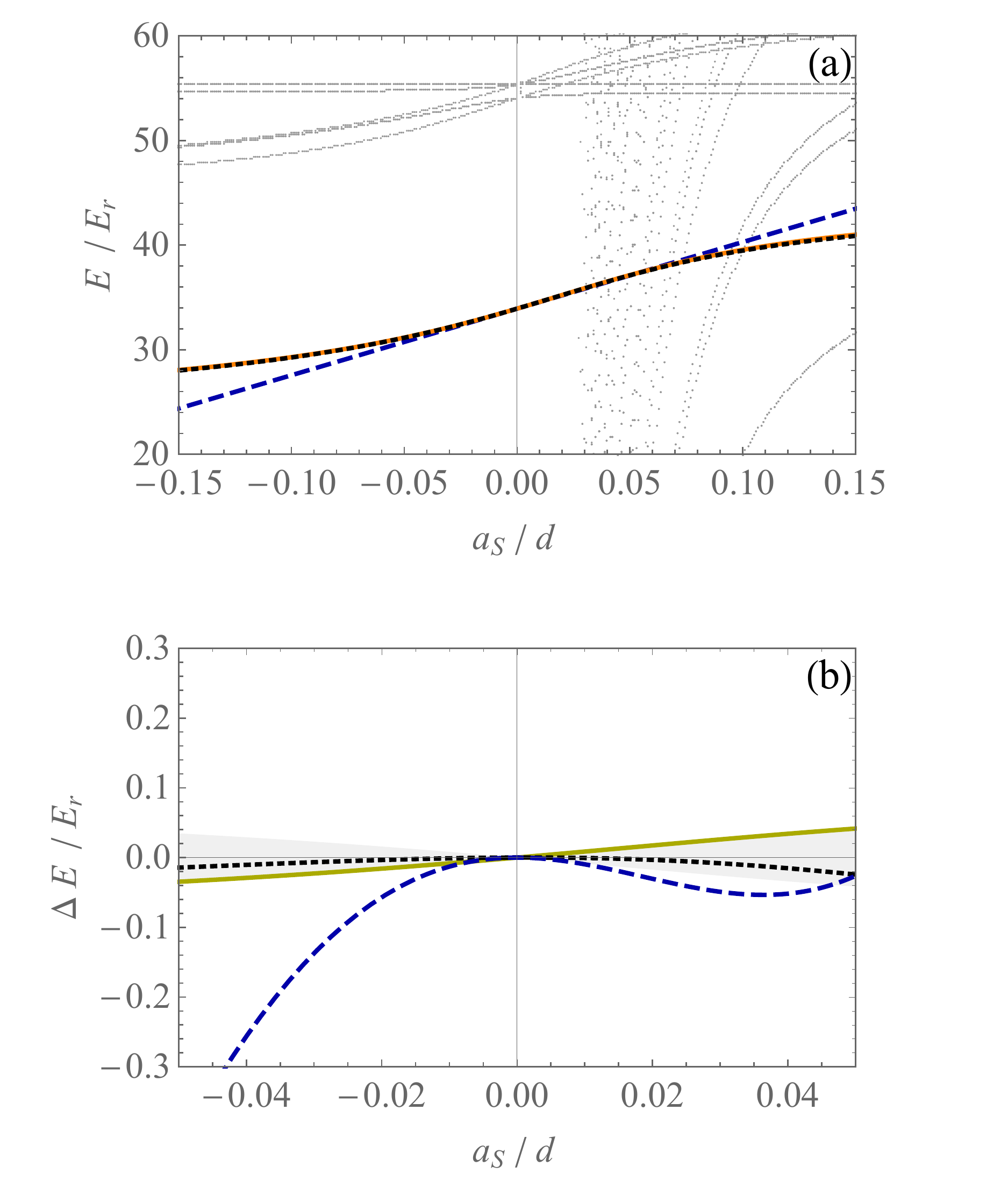}
\caption{\label{fig:fig1} Comparison of the full contact system (gray dots, black dotted line) to the linear model (blue dashed line) and the effective harmonic model (orange full line) for a lattice depth of $V_0 = 35 E_\text{r}$.
(a) Energy spectrum as a function of the scattering length $a_S$. The black dotted line describes the energy of the state connected to the lowest band Wannier state at $a_S = 0$, the gray dots crossing it are related to the deeply bound dimer state in different excited center of mass bands, while the remaining gray dots are connected to excited bands. 
(b) Energy spectrum relative to the effective harmonic model. Linear model (blue dashed), the full contact model (black dotted) or the effective harmonic model determined from the systems Wannier function including side peaks (yellow full). The gray shaded array indicates our estimate of the models theoretical error, which we estimate to be the maximum of the deviation to the full numerical and the effective harmonic model determined from the systems Wannier function (black dotted and yellow full lines). }
\end{figure}

We want to determine the spectrum of the full Hamiltonian close to the ground state energy of the non-interacting full system. 
Therefore we project the full Hamiltonian on the states $|N_X, N_Y, N_Z, n, \ell, |m| \rangle_{S/A}$. We cut the number of base states by restricting to states with $E(N_X, N_Y, N_Z, n, \ell, m) < E_\text{max}$ or a finite range of the quantum numbers. 
For each such sub-base we also restrict to a finite expansion of $v_\Delta (X,x) \approx \sum_{i,j}^{max} \alpha_{ij}(V_0,V_\text{eff}) x^{2 i}X^{2 j}$. We chose to expand up to $10$th order in $2i + 2j$, leading to a good approximation of the lattice potential around a single lattice site up to $\pm 0.7 d$, with $d$ the lattice constant.
Diagonalizing the full Hamiltonian projected on the sub-base we can obtain a converged spectrum around the energy of the lowest band. In Fig.~\ref{fig:fig1} we compare the energy shifts $U$ from the  first order perturbative and the effective description to the full solution for an optical lattice of depth $V_0 = 35 E_\text{r}$.
We assume here a lattice constant $d$ of $532 \text{nm} = 10053 a_0$, but the result will depend only on the ratio $a_S / d$. 
We get good agreement with the full result for the effective harmonic model with a maximal relative deviation of about $<0.8\%$ for $|a_S| < 0.05 d \approx 500 a_0 $.
Note that the results presented here are just valid around a single lattice site $\pm 0.7 d$, therefore the side peaks of the Wannier function cannot be correctly accounted for. 
We estimate the error caused by this deviation by comparing the effective harmonic models determined with and without including the peaks at neighboring lattice sites (cf. Fig.~\ref{fig:fig1} (b)), which leads us to an estimate of the theoretical accuracy of our model indicated by the gray shaded area in Fig.~\ref{fig:fig1}. 
With that we still have a maximal relative systematic error in $U$ of $1.3\%$ for $|a_S| < 0.05 d$.

When interpreting experimental data on $U$ the refined harmonic description leads to corrections in the position and width of Feshbach resonances as compared to the linear model. With the refined model one finds resonance positions shifted towards the direction of negative scattering lengths as well as increased resonance widths. The shifts in resonance positions result from non anti-symmetric behavior of $U$, while the increase in width is related to the flattening off of $U$ for diverging $a_S$.

\subsection{Connection to a coupled-channels model}

So far we considered particles interacting via a contact potential.
In ultracold gas experiments the collisions are usually considered to happen at zero energy and thus the scattering lengths of the coupled-channels and contact systems need to be matched. 
For the considered lattice geometry however we can estimate the relative energy at which the particles collide to be around the energy $\epsilon$ of the lowest single particle band, which is for our parameters $\epsilon / k_B \approx 20\mu K$.
This is estimated from two non-interacting particles in the lowest band with energy $2 \epsilon$. Approximately half of this is relative energy, which is also the collision energy at zero separation. Adding a repulsive interaction leads to an increase in collision energy since the wave function needs to shift to higher energies, which also leads to an increase in size of the wave function in relative direction.
Despite the wider spatial and therefore narrower momentum distribution at low momenta of the two-particle wave function, there is an increase in average relative kinetic energy due to an increase in the high momentum tail.

In general
a contact model can be adjusted to correctly represent the s-wave component of any short range potential for far separation of the particles at a fixed collision energy $\epsilon = \hbar^2 k^2 / (2 \mu )$.
To do so one has to choose the following scattering length for the contact system
\begin{equation} \label{eq:matchcond}
 \frac{1}{a_S} = - k \text{cot} (\delta^{\text{cc}}(k)) \, ,
\end{equation}
with $\delta^{\text{cc}}(k)$ the scattering phase shift of the full coupled-channels interaction.
The right hand side can be expanded in powers of $k$ with
\begin{equation}
k \text{cot} (\delta^\text{cc}(k)) 
= - \frac{1}{a_S^\text{cc}} + \frac{R_e}{2} k^2 + \mathcal{O} \left( k^4 \right) \, ,
\end{equation}
with $a_S^\text{cc}$ the scattering length of the coupled-channels model and $R_e$ the effective range.
This leads us to the implications of including collisions at a finite energy.
Effects due to finite collision energy $\epsilon$ are relevant especially close to narrow resonances, where the effective range $R_e$ can take large values. For narrow resonances the resonance position at finite collision energy $B_0(\epsilon)$ defined by 
\begin{equation}
k \text{cot} (\delta^\text{cc}(B_0,k)) 
= 0 
\end{equation}
can thus be subject to shifts of order $\Delta B_0 (\epsilon)\sim \epsilon / \mu_\text{B} \approx 0.3 $G.  

Similar to the free case Eq.~(\ref{eq:matchcond}) we want to arrive at a  matching condition for the lattice environment. 
Since the lattice system cannot be reduced to relative and c.m. motion there will be no well defined collision energy.
Instead the collision is happening in different c.m. channels simultaneously, where each of those channels has an assigned collision energy. 
To see this we split the Hamiltonian into three components
\begin{align}
H & = \frac{\mathbf{P}^2}{2 m} 
+ \frac{\mathbf{p}^2}{2 m} 
+ V_\text{int} (r)
+ V_\text{opt}(\mathbf{r},\mathbf{R}) \\
& = \frac{\mathbf{P}^2}{2 m} 
+ V_\text{opt}(0,\mathbf{R}) \nonumber \\
& \phantom{ = \frac{\mathbf{P}^2}{2 m} } + \frac{\mathbf{p}^2}{2 m} 
+ V_\text{int} (r)
+ \tilde{V}_\Delta(\mathbf{r} ,\mathbf{R})\\
& = H_{\mathbf{R}} + H_{\mathbf{r}} + \tilde{V}_\Delta(\mathbf{r} ,\mathbf{R}) 
\, ,
\end{align}
one acting solely on the c.m. component $H_{\mathbf{R}}$, one acting solely on the relative coordinate $H_{\mathbf{r}}$, and a part acting on both $\tilde{V}_\Delta(\mathbf{r} ,\mathbf{R})$.

Here $H_\mathbf{R}$ describes a lattice system in the c.m. coordinate. 
Note however that the depth of the trapping potential is twice as big as in the cartesian directions, while the lattice constant is bigger by a factor of $\sqrt{2}$, which in total leads to a lattice that is effectively $4$ times deeper than the original one (cf. Eq.~(\ref{eq:PotCOM})). 
We can get an approximate expression for the full Hamiltonian valid around a single lattice site by projecting on the c.m. Wannier functions $| i \rangle$
at that given site with associated c.m. band energies $\mathcal{E}_i$.
\begin{align}
H & \approx \sum_{i,i'} | i \rangle \langle i' |\left[\delta_{ii'}(H_\mathbf{r} + \mathcal{E}_i) + \langle i | \tilde{V}_\Delta | i' \rangle \right] 
\, .
\end{align}
In addition we find that the coupling term $\tilde{V}_\Delta$ vanishes up to second order in $r$ as $r \rightarrow 0$. For clarity we give the terms $V_\text{opt}(0,\mathbf{R})$ and $\tilde{V}_\Delta(\mathbf{r} ,\mathbf{R})$ explicitly
\begin{widetext}
\begin{align}
V_\text{opt}(0,\mathbf{R}) & = 2 E_\text{r} V_0 \left[ 
\text{sin}(k_\text{r} X / \sqrt{2} )^2 
+ \text{sin}(k_\text{r} Y / \sqrt{2} )^2
+ \text{sin}(k_\text{r} Z / \sqrt{2} )^2
\right] \label{eq:PotCOM} \\
\tilde{V}_\Delta(\mathbf{r} ,\mathbf{R}) & =
E_\text{r} V_0 \left[ (1 - \text{cos}(\sqrt{2} k_\text{r} x )) 
\text{cos}[\sqrt{2} k_\text{r} X]
+( 1 - \text{cos}(\sqrt{2} k_\text{r} y )) \text{cos}(\sqrt{2} k_\text{r} Y) \right. \\
& \phantom{=} \left. \qquad \qquad \qquad 
+ ( 1 - \text{cos}(\sqrt{2} k_\text{r} z ))
\text{cos}(\sqrt{2} k_\text{r} Z)\right] \nonumber \\
& \approx (k_\text{r} r)^2 \tilde{V}_\Delta^{(2)}(\hat{\mathbf{r}} ,\mathbf{R}) + \mathcal{O}((k_\text{r}r)^4) \, .
\end{align}
\end{widetext}
For a contact interaction the inner boundary condition in all channels is determined by the scattering length. 
Upon investigating the full numerical contact solutions $\Psi$ we can get a lower bound for the population of the lowest c.m. channel $| 0 \rangle$ by
\begin{align}
& |\langle \Psi|\left(| 0 \rangle\langle 0 |\otimes \mathbb{1}_\mathbf{r} \right) |\Psi \rangle| \nonumber \\
& \geq \alpha^2 \beta^2 - 2 \alpha \beta \sqrt{1 - \alpha^2}\sqrt{1 - \beta^2} \, ,
\end{align}
with $ \alpha = | \langle 0 | \mathbf{0} \rangle|$ and $\beta = |\langle \Psi|\left(| \mathbf{0} \rangle\langle \mathbf{0} |\otimes \mathbb{1}_\mathbf{r} \right) |\Psi \rangle|$, where $| \mathbf{0} \rangle = | N_X = 0, N_Y = 0, N_Z = 0 \rangle$ denotes the lowest harmonic oscillator state in the c.m. component.
For the bound state considered we find that the lowest c.m. channel is always populated to more than $99\%$ in the considered scattering length regime.
Therefore we expect to get a good approximation by matching the contact solution in this channel to the full coupled-channels result.
The strength of the contact interaction and thus the scattering length $a_S$ can be determined from a coupled-channels model by justifying the condition (cf. Eq.~(\ref{eq:ueffharm}))
\begin{equation} \label{eq:matchcc}
\frac{\sqrt{2} \,  \Gamma \left(\frac{U(k)}{2  E_\text{HO}} \right)}{\ell_\text{HO} \Gamma \left(\frac{U(k)}{2  E_\text{HO}} - \frac{1}{2}\right)}
= - k \text{cot} (\delta^\text{cc}(k))  \, ,
\end{equation}
where the on-site interaction energy $U(k)$ is determined by $\hbar^2 k^2 / (2 \mu ) = U(k) + E(0) - \mathcal{E}_0$ with $\mathcal{E}_0$ the energy of the lowest c.m. band.
This ensures that the full contact and the coupled-channels wave function are properly matched in the lowest c.m. band. 
The boundary condition in the other populated collision channels is however not exactly satisfied this thus limits the contact model approach to regimes where either the effective range correction terms are small or the lowest c.m. channel is dominating.

To sum up in Eq.~(\ref{eq:matchcc}) we combine an effective harmonic parametrization of $U$ representing a contact interaction model with the phase shift behavior $\delta^\text{cc}(k)$ of a full coupled-channels calculation. This enables us to perform a coupled-channels analysis of on site interaction data \citep{Amato-Grill:2019} in the following section.

\section{comparison to experiment}

Our new model is captured by Eq.~(\ref{eq:matchcc}). We apply it to recent experimental data taken for $^7$Li~\cite{Amato-Grill:2019}.
For $^6$Li and $^7$Li the ultracold interatomic interactions already have been accurately characterized by different experiments, however, also discrepancies are known to exist directly related to the two-body interactions, for instance in the determination of three-body parameters near an $|f,m_f\rangle=|1,1\rangle$ Feshbach resonance of $^7$Li \cite{Dyke:2013,Gross:2011}. 
In~\cite{Amato-Grill:2019} a sample of ultra-cold $^7$Li atoms is prepared in a Mott insulating state with doubly-occupied sites in a cubic optical lattice with a dept of $35 E_r$.
The experiment involves the two lowest hyperfine states labeled as 
$
 | a \rangle$ and $
| b \rangle$. More precisely, at each doubly-occupied lattice site one of the three symmetric combinations in spin $| a a \rangle$, $|a b \rangle_S = (| a b \rangle + | b a \rangle)/\sqrt{2}$ or $| b b \rangle$ can be realized and correspond to one of three different interaction channels.
With the help of radio frequency pulses transitions between the different spin states can be driven. From the resonance frequency positions the difference in on-site interaction energy between the scattering channels $(U_{ab} - U_{aa})$ and $(U_{bb} - U_{ab})$ can be inferred.
This can be done for a wide range of external magnetic fields $B$.

We note that the analysis presented here relies on the effective harmonic model using the oscillator frequency determined from the numerical model (orange line in Fig.~\ref{fig:fig1} (a)). 
We include the gray shaded area in Fig.~\ref{fig:fig1} (b) as our theoretical error estimate.

\begin{figure*}[htb!] 
	  \centering
  \begin{minipage}[c]{0.45\textwidth}
    \includegraphics[width=\textwidth]{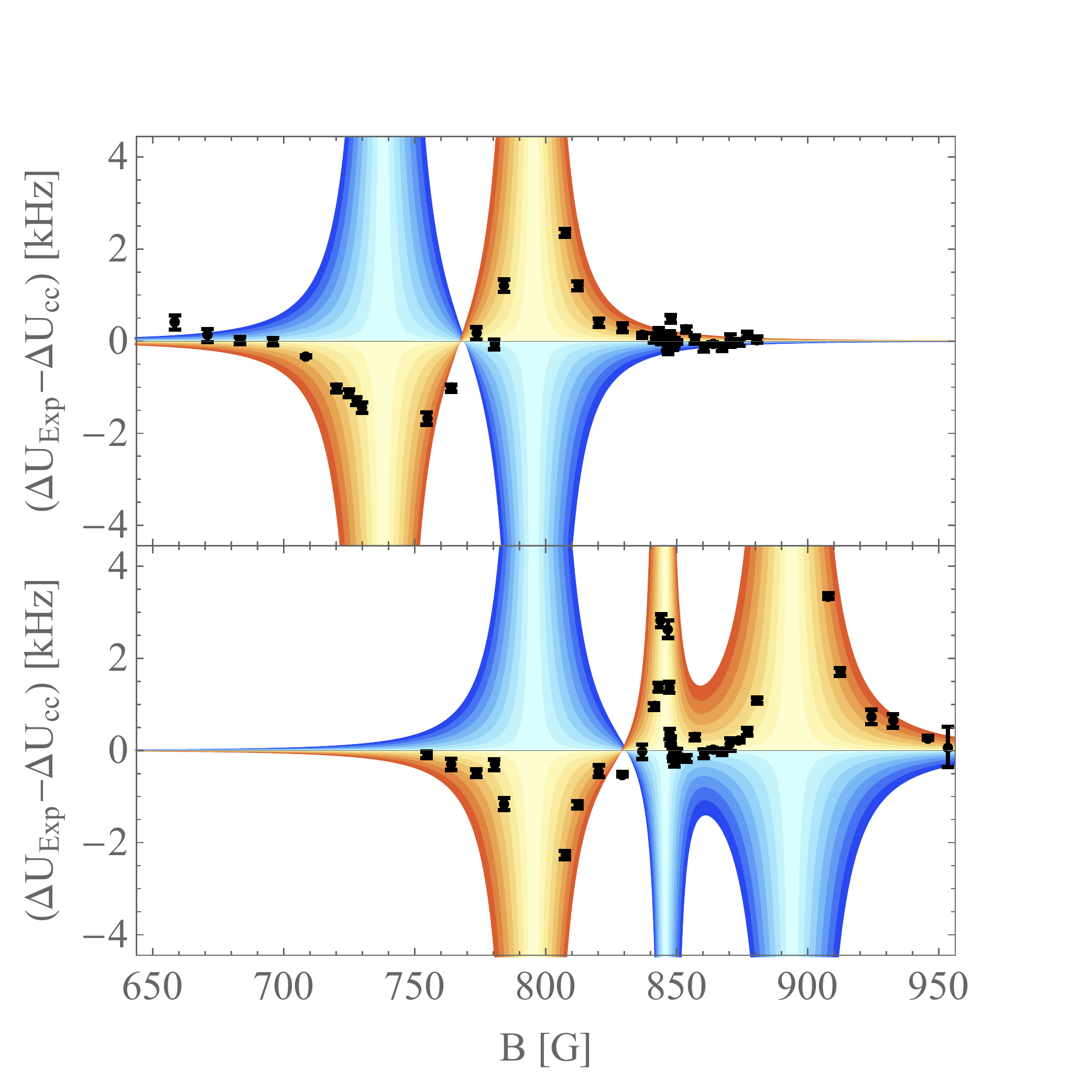}
  \end{minipage}
  \centering
  \begin{minipage}[c]{0.45\textwidth}
    \includegraphics[width=\textwidth]{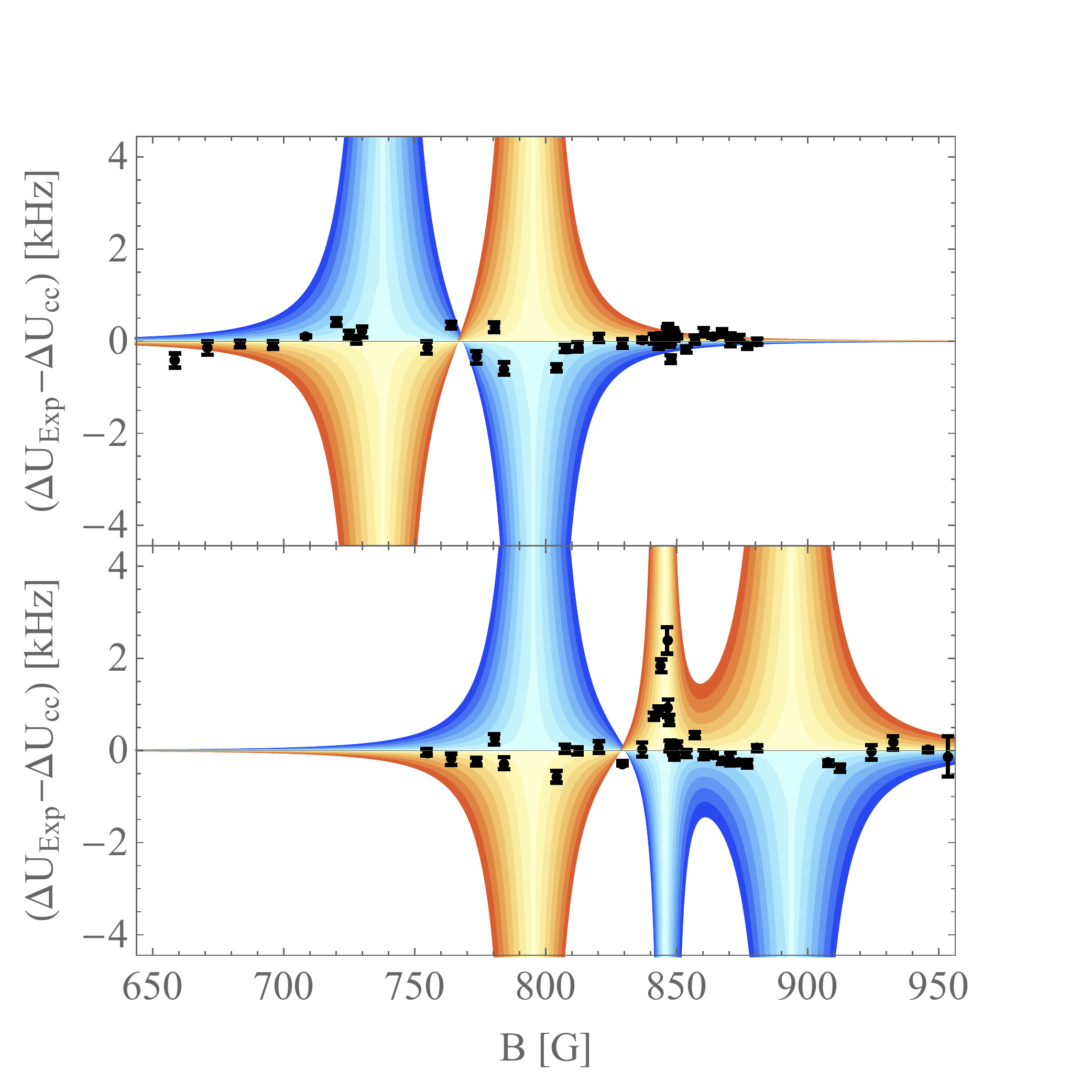}
  \end{minipage}
  \centering
  \begin{minipage}[c]{0.07\textwidth}
    \includegraphics[width=\textwidth]{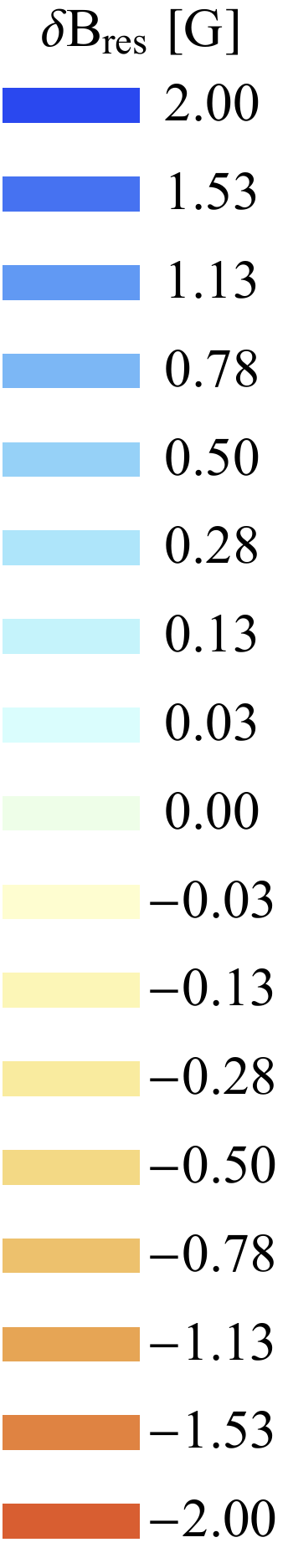}
  \end{minipage}
\caption{\label{fig:fig4} Comparison of the experimental to the coupled-channels model determined in~\cite{Gross:2011}. We show the residuals between experimental and theoretical data for the $| aa \rangle$ to $| ab \rangle$ transition (top) and the $| bb \rangle$ to $| ab \rangle$ transition (bottom). The plots on the left hand side have been obtained with the linear the ones on the right hand side with the effective harmonic model.
The color scale indicates the shift $\delta B_\text{res}$ required in the theoretical resonance positions to reach agreement. 
Thus the color scale can serve as a simple measure for the quality of agreement between the unfitted coupled-channels model and the experimental data. 
The color map has been obtained starting from a dispersive model fitted to the coupled-channels result.
}
\end{figure*}

As a first test we want to verify that the effective model indeed leads to an improvement compared to the linear Hubbard description. For that purpose, we compare the experimental interaction spectroscopic data to a coupled-channels model gauged to earlier experiments~\cite{Gross:2011}. 
We use the scattering length data determined from this
model and map it with either the linear Eq.~(\ref{eq:linconv}) or the effective harmonic description Eq.~(\ref{eq:ueffharm}) onto the experimental data. 
In Fig.~\ref{fig:fig4} we show that the effective harmonic model leads to improved agreement with the experimental data. 
There we compare the differences in on site interaction energy $\Delta U$ by showing the deviations $\Delta U_\text{exp} - \Delta U_\text{cc}$ between experiment and theory for better visibility. We find as expected that the resonance positions in $\Delta U$ for the linear conversion appear at too high magnetic field values.

\begin{figure*}[hbt!]
	  \centering
    \includegraphics[width=\textwidth]{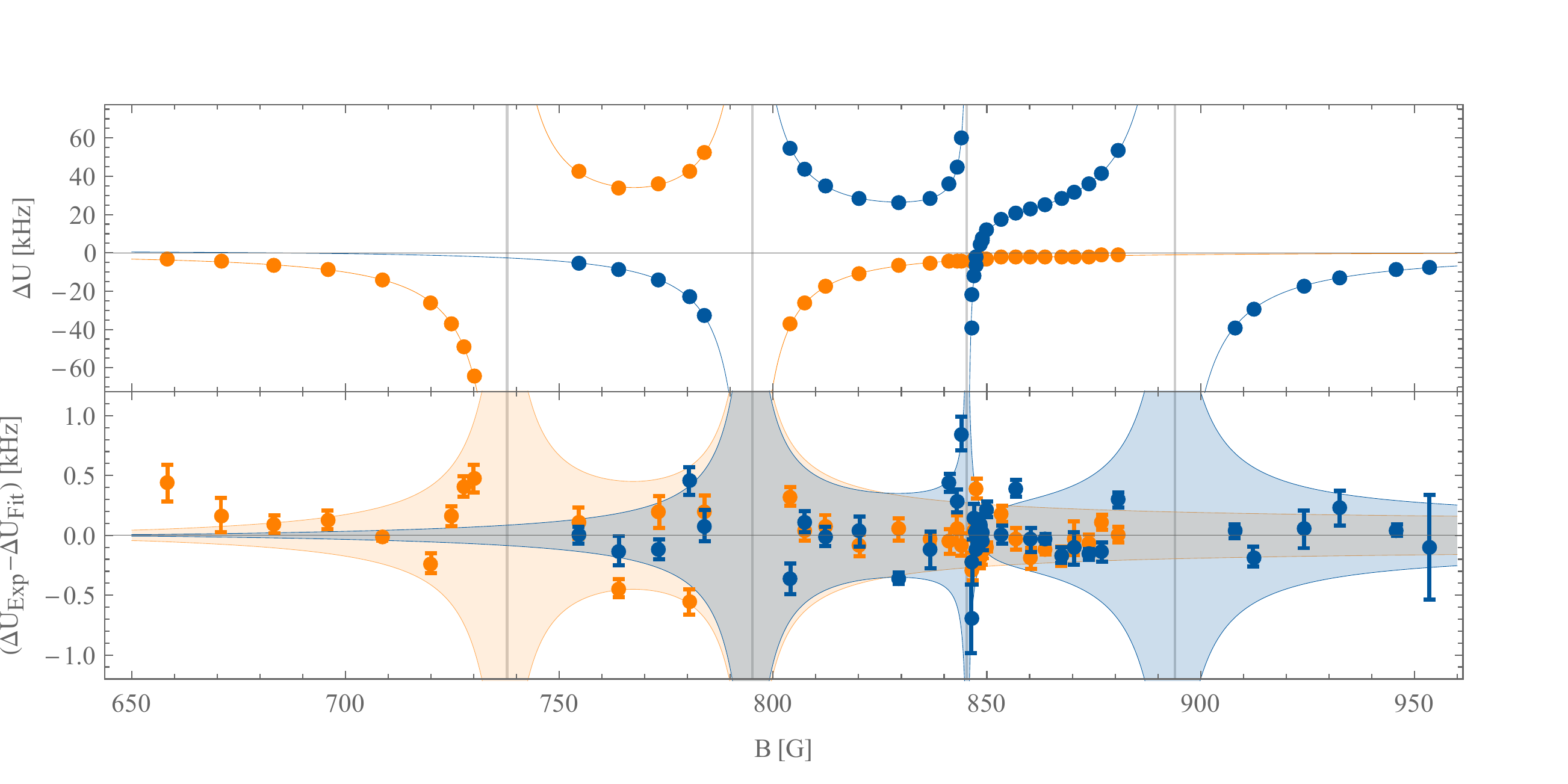}
\caption{\label{fig:fig3} Experimental data $U_{ab} - U_{aa}$ (orange points) and $U_{bb} - U_{ab}$ (blue points) and best fit with the coupled-channels description to the combined data set (blue and red lines), the lower half shows the difference of the data points and the fitted model along with the experimental error bars taken to be one $\sigma$ of the fit to the resonance spectra, while the systematic error on the experimental side has been estimated to  $0.1$kHz. Our estimate of the theoretical uncertainties is indicated by the orange and blue shaded areas.
}
\end{figure*}

We use the effective harmonic approximation including finite collision energy effects Eq.~(\ref{eq:matchcc}) to map coupled-channels phase shift data $\delta^\text{cc}$ to on-site interaction energy $U$. 
We take the coupled-channels model presented in detail in~\cite{Gross:2011}, where it was used to interpret rf-spectroscopy data taken for $^7$Li.
The most crucial parameters in the coupled-channels model are the van der Waals coefficient $C_6$ and the adjustments in the singlet $S$ and triplet $T$ boundary conditions parametrized in form of phase parameters $\Delta \phi_S$ and $\Delta \phi_T$~\cite{Gross:2011}. We take those as free parameters that we fit to the experimental data by performing a $\chi^2$ minimization.
Our fit results are presented in Fig.~\ref{fig:fig3} and Tab.~\ref{tab:tab1}.
The theoretical error estimate has been included into our error analysis by adding the theoretical error determined from the model of Ref.~\cite{Gross:2011} to the experimental error bars prior to the fit.
We find good agreement with the experimental data (cf. Fig.~\ref{fig:fig3}), just close to the narrow resonance our model seems to underestimate the width of the resonance. 
Note that we identified the regime close to the narrow Feshbach resonances to be less well approximated by the contact model.
In Tab.~\ref{tab:tab1} we compare to the results obtained in~\cite{Amato-Grill:2019} with a linear Hubbard model in combination with dispersive shapes to parametrize $a_S$ in the different interaction channels. For the positions of the broad resonances (at $738$G, $795$G and $894$G) we find values corrected by $0.2$ to $0.7$G to higher magnetic fields. 
We attribute the major contribution to these corrections to the effective harmonic model. 
However, for the narrow resonance (at $845$G) we find a correction to lower magnetic fields of $0.1$G as a result of the finite collision energy effects included.
Comparing to previous results we find improved agreement for the broad resonances, while the deviation in the position of the narrow resonance increased.

Similar good agreement could be achieved by fitting with the effective harmonic model without finite collision energy corrections.
The resulting resonance positions for the broad resonances are in agreement within the two different models while the narrow resonance is shifted to lower values for the energy corrected model.

\begin{table*}[htb!]
\centering 
\begin{tabular*}{\textwidth}{@{\extracolsep{\fill}}lccc@{}}
\hline
\hline
channel & $B_\text{res}$ [G] & $B_\text{res}$ [G]  & $B_\text{res}$ [G] \\
 & (coupled-channels fit) & (taken from \cite{Amato-Grill:2019})  & (previous works)\\
\hline
aa & 737.81(2) & 737.58(10) & 737.88(2) \cite{Gross:2011} \\
&&& 737.8(2) \cite{Navon:2011}\\
&&& 737.69(12) \cite{Dyke:2013}\\
ab & 795.20(2) & 794.64(07) & \\
bb & 845.322(5) & 845.42(01) & 845.54  \cite{Gross:2011} \footnote{There is no error bar given in \cite{Gross:2011}}\\
bb & 894.00(4) & 893.34(12) & 893.95(5) \cite{Gross:2011} \\
\hline
\hline
\end{tabular*}
\caption{\label{tab:tab1} Resonance positions determined from the coupled-channels code compared to the analysis of the same data done in \cite{Amato-Grill:2019} using the linear Hubbard model in combination with dispersive shapes to parametrize $a_S$ in the different interaction channels and to previous works. Our error bars are taken to be one standard deviation.}
\end{table*}

\section{conclusion and outlook}

We presented a full coupled-channels description of the on-site interaction energy $U$ of a Mott insulator state  with two atoms per lattice site. 
Our description is based on a parametrization of the on-site interaction energy with an effective harmonic model adjusted to match the linear behavior at small interaction strengths that can be determined from the Hubbard model.
A matching condition has been obtained Eq.~(\ref{eq:matchcc}) that combines the effective harmonic parametrization of $U$ with the relative collision energy in the lowest c.m. band from the non-interacting scenario.
We verified that for moderate scattering lengths up to $0.05 d$ and a lattice dept $V_0 = 35 E_\text{r}$ our effective approach gives good agreement with the full contact scenario.
We applied our effective description to perform a successful coupled-channels analysis of recent experimental data on $^7$Li~\cite{Amato-Grill:2019}.
The high precision of the data enabled us to demonstrate that our effective harmonic description is an improvement to the linear Hubbard model.
We show that including finite collision energy effects leads to a change in the predicted resonance positions especially for narrow resonances.  
The analysis of the experimental data however did not allow to undermine an improvement from the model without to the model with finite collision energy effects.
Still a precise independent determination of the resonance position especially for the narrow resonance in the $|bb \rangle$ channel could easily lead to such a distinction.
Overall we find that our model is in good agreement with the experimental data, but close to the narrow resonance our model seems to slightly underestimate the width of the resonance. 
Note that we identified the regime close to narrow Feshbach resonances to be the one least well approximated by the contact model.
However a refined description valid also close to narrow resonances could be obtained by matching the effective harmonic model to the coupled-channels model for each base state involved before the direct diagonalization method is applied, or by introducing the lattice potential directly into the coupled-channels calculation, as it has been done for the single channel case~\cite{Grishkevich:2009,Grishkevich:2011}. 
The energy matching condition could thus be satisfied exactly in those cases, but the advantage of having a single matching condition would be lost.

Measurements of the on site interaction energy shift for different lattice depths could reveal the dependence of the resonance positions as a function of collision energy. Also note that the points in magnetic field, where the scattering lengths of two channels are equal should be independent of the conversion model, in a regime where the scattering lengths determines the interaction. 
These are the points where $(U_{ab} - U_{aa})$ or $(U_{bb} - U_{ab})$ cross zero or each other. A precise determination of those points might be valuable information in addition to the resonance positions.

\section*{Acknowledgements}
We thank Paul Mestrom, Victor Colussi, Wouter Verhoeven, Silvia Musolino, Denise Braun and Jinglun Li for fruitful discussions.
This research is financially supported by the Netherlands Organisation for Scientific Research (NWO) under Grant 680-47-623, and is part of the research programme of the Foundation for Fundamental Research on Matter (FOM). 
We also acknowledge support from the NSF through the Center for Ultracold Atoms and Award No. 1506369, from ARO-MURI Non-Equilibrium Many-Body Dynamics (Grant No. W911NF14-1-0003), from AFOSR-MURI Quantum Phases of Matter (Grant No. FA9550-14-10035), from ONR (Grant No.N00014-17-1-2253), and from a Vannevar-Bush Faculty Fellowship. J.A-G. acknowledges support from the National Science Foundation Graduate Research Fellowship under GrantNo. DGE 1144152.

\bibliography{biblio-TS}

\end{document}